\documentclass[twocolumn,showpacs]{revtex4}
\usepackage[dvips]{graphicx}
\usepackage{amsmath,amssymb,bm}

 \def\a{\alpha}

\def\be{\begin{equation}}
\def\ee{\end{equation}}
\def\bea{\begin{eqnarray}}
\def\eea{\end{eqnarray}}

\def\Lag{{\mathcal{L}}}

\begin{document}

\title{Holographic conductivity of zero temperature superconductors}
\author{R. A. Konoplya}\email{konoplya_roma@yahoo.com}
\affiliation{Department of Physics, Kyoto University, Kyoto 606-8501, Japan\\
and\\
Theoretical Astrophysics, Eberhard-Karls University of
T\"{u}bingen, T\"{u}bingen 72076, Germany}
\author{A. Zhidenko}\email{zhidenko@fma.if.usp.br}
\affiliation{Instituto de F\'{\i}sica, Universidade de S\~{a}o Paulo\\ C.P. 66318, 05315-970, S\~{a}o Paulo-SP, Brazil}

\begin{abstract}
Using the recently found by G. Horowitz and M. Roberts
(arXiv:0908.3677) numerical model of the ground state of
holographic superconductors (at zero temperature), we calculate
the conductivity for such models. The universal relation
connecting conductivity with the reflection coefficient was used
for finding the conductivity by the WKB approach. The dependence
of the conductivity on the frequency and charge density is
discussed. Numerical calculations confirm the general arguments of
(arXiv:0908.3677) in favor of non-zero conductivity even at zero
temperature. In addition to the Horowitz-Roberts solution we have
found (probably infinite) set of extra solutions which are
normalizable and reach the same correct RN-AdS asymptotic at
spatial infinity. These extra solutions (which correspond to
larger values of the grand canonical potential) lead to effective
potentials that also vanish at the horizon and thus correspond to
a non-zero conductivity at zero temperature.
\end{abstract}
\maketitle

\section{Introduction}

The famous AdS/CFT correspondence \cite{AdS/CFT} allows to
describe conformal field theory in $d$-dimensional space-time by
considering a $d+1$-dimensional super-gravity in anti-de Sitter
space-time. This opens a number of opportunities to look into
non-perturbative quantum field theory at strong coupling. One of
the recent interesting applications of such a holography is
constructing of a model of a superconductor.  Usually in quantum
field theory superconductors are well understood by the
Bardeen-Cooper-Schrieffer, theory \cite{Bardeen:1957mv}, though
there are indications that for some systems the standard Fermi
liquid theory cannot be a good approximation
\cite{superconductor0}. Therefore a holographic model for
superconductors was suggested by Hartnoll, Herzog and Horowitz
\cite{superconductor1}. This model have been recently studied in a
number of papers and some alternative models of holographic
superconductors were suggested
\cite{Brynjolfsson:2009ct}-\cite{Minic:2008an}. These models
contain a charged asymptotically anti-de Sitter black hole which
have non-trivial hairs at low temperatures. Until recent time,
there were suggested various holographic models for the low
temperature limit, while the dual description for the actual zero
temperature ground state remained unknown. The very recent paper
of Horowitz and Roberts \cite{Roberts} solves this problem and
find numerically the zero temperature holographic dual for
superconductors.

The system under consideration consists of the charged scalar
field coupled to a charged ($3+1$)-dimensional black hole, so that
above some critical temperature, in the normal phase, the system
is described by the Reissner-Nordstr\"om-anti-de Sitter black
holes, while below the critical temperature, in the
super-conducting phase, the black hole develops scalar hairs.
Thus, the superconductor is ($2+1$)-dimensional, what might be
realized for instance in graphene. In \cite{Roberts}, based on
qualitative arguments, it has been shown that the effective
potential of the perturbation equation for the dynamic of the
Maxwell field vanishes at the horizon, and, consequently, the
conductivity never vanishes even at zero temperature. Though some
intuitive arguments were given in \cite{Roberts} about the
behavior of conductivity in the suggested model, no calculations
of conductivity were performed there, except for some estimations
made for the low-frequency regime. Therefore our first aim here
was to calculate conductivity for the Horowitz-Roberts model
\cite{Roberts}. When integrating the field equations, in addition
to the ground state solution described in \cite{Roberts}, we have
found  a number of other solutions with the same leading AdS
asymptotic at spatial infinity and obeying the same general form
of anzats near the horizon. We have checked that the found here
extra solutions, as that of \cite{Roberts}, have vanishing
effective potential at the horizon, so that the conductivity will
never be zero even at zero temperature. They correspond to
configurations of the scalar field with higher energies at zero
temperature.

The paper is organized as follows. Sec II gives the basics
equations for the system of fields under consideration and scheme
of construction of the numerical solution for a black hole with
the scalar hair. Sec III describes the spectrum of the obtained
solutions which consists of the ground state solution and
solutions of the higher grand canonical potential. Sec IV is
devoted to WKB calculations of the conductivity for the
zero-temperature superconductor.

\section{Construction of the Horowitz-Roberts holographic dual}

The Lagrangian density for the system under consideration takes the form
\begin{equation}\label{eq:bulktheory}
\Lag = R + \frac{6}{L^2} - \frac{1}{4} F^{\mu\nu} F_{\mu\nu}
- |\nabla \psi - i q A \psi |^2 -U( |\psi|) \,
\end{equation}
where $\psi$ is the scalar field, $F_{\mu\nu}$ is the strength tensor of electromagnetic field, $m,q$ are the scalar field`s charge and mass and $A$ is the vector-potential ($F=dA$). The cosmological constant is $-3/L^2$.
The plane symmetric solution can written in a general form
\be\label{metric}
 ds^2=-g(r) e^{-\chi(r)} dt^2+{dr^2\over g(r)}+r^2(dx^2+dy^2)
\end{equation}
\begin{equation}
A=\phi(r)~dt, \quad \psi = \psi(r)
\end{equation}
We shall fix the gauge so that $\psi$ is real and measure all the
quantities in units of the AdS radius, so that $L=1$. The
equations have the form:
\begin{subequations}\label{nlem}
\begin{equation}
 \psi''+\left(\frac{g'}{g}-\frac{\chi'}{2}+\frac{2}{r} \right)\psi' +\frac{q^2\phi^2e^\chi}{g^2}  \psi  -{U'(\psi)\over 2g}=0\label{psieom}
\end{equation}

\begin{equation}\label{phieom}
\phi''+\left(\frac{\chi'}{2}+\frac{2}{r}  \right)\phi'-\frac{2q^2\psi^2}{g}\phi=0
\end{equation}

\begin{equation}
\chi'+r\psi'^2+\frac{rq^2\phi^2\psi^2e^\chi}{g^2}=0\label{chieom}
\end{equation}

\begin{equation}\label{geom}
g' + \left(\frac{1}{r}  - { \chi'\over 2}\right)
g+\frac{r\phi'^2e^\chi}{4}- 3r+\frac{rU(\psi)}{2}=0
\end{equation}
\end{subequations}

When we choose $\chi =0$ at infinity, the metric takes the standard AdS form at larger $r$.
\begin{equation}
\phi = \mu -{\rho\over r}, \qquad \psi ={\psi^{(\lambda)}\over r^\lambda}+{\psi^{(3-\lambda)}\over r^{3-\lambda}}.
\end{equation}
where $\lambda = (3 +\sqrt{9+4m^2})/2 $. In the boundary dual CFT, $\mu$ is the chemical potential, $\rho$ is the charge density, and $\lambda$ is the scaling dimension of the operator dual to $\psi$. We used
\begin{equation}\label{normcond}
\psi^{(3-\lambda)} = 0.
\end{equation}

The density of the grand canonical potential $\Omega$ of the state
that corresponds to a given solution can be find by fitting the
function $g(r)$ at large r \cite{Hartnoll:2008kx}
\begin{equation}\label{gpfit}
e^{-\chi(r)}g(r)=r^2+\frac{\Omega}{r}+o\left(\frac{1}{r}\right).
\end{equation}

Let us consider two cases:
\begin{enumerate}
\item
The case $m^2 = 0$ corresponds to a marginal operator, $\lambda
=3$, in the $2+1$ superconductor with a nonzero expectation value.
Following \cite{Roberts} we have used the ansatz
\begin{eqnarray}\label{ansatz}
\phi = r^{2+\a},&& \psi = \psi_0 - \psi_1 r^{2(1+\a)},
\\\nonumber
\chi =\chi_0 - \chi_1 r^{2(1+\a)}, && g=r^2(1 - g_1r^{2(1+\a)}).
\end{eqnarray}
The coefficients in $\phi$ and $g$ can be taken equal to unity.
Substituting this into the field equations  and equating the
dominant terms for small $r$ (with $\a > -1$), one has
\begin{eqnarray}
q\psi_0 &=& \left({\a^2 + 5\a + 6\over 2}\right)^{1/2}, \nonumber\\
\chi_1 &=& {\a^2 + 5\a + 6\over 4(\a + 1)}e^{\chi_0},\quad
 g_1 =  {\a
+ 2\over 4} e^{\chi_0}, \\\nonumber \psi_1 &=& {q e^{\chi_0}
\over 2(2\a^2 + 7\a +5)}\left({\a^2 + 5\a + 6\over
2}\right)^{1/2}.
\end{eqnarray}
These formulas were obtained in \cite{Roberts}.

We solve the equations (\ref{nlem}) numerically using the ansatz
(\ref{ansatz}). We choose $\alpha$ in order to satisfy the
condition (\ref{normcond}) using the shooting algorithm.

\item
In \cite{Roberts} the ansatz for small $r$ has been found for the
case of $m^2<0$ and $q^2>-m^2/6$ ($\lambda<3$):
\begin{eqnarray}\label{ans}
\phi = \phi_0r^{\beta}\left(-\ln(r)\right)^{1/2},&& \psi =
2\left(-\ln(r)\right)^{1/2},
\\\nonumber
\chi =\chi_0 + \ln(-\ln(r)), && g=(2m^2/3)r^2\ln(r),
\end{eqnarray}
where
$$\beta=-\frac{1}{2}+\sqrt{1-\frac{48q^2}{m^2}}>1.$$

\begin{figure}
\centerline{\resizebox{\linewidth}{!}{\includegraphics*{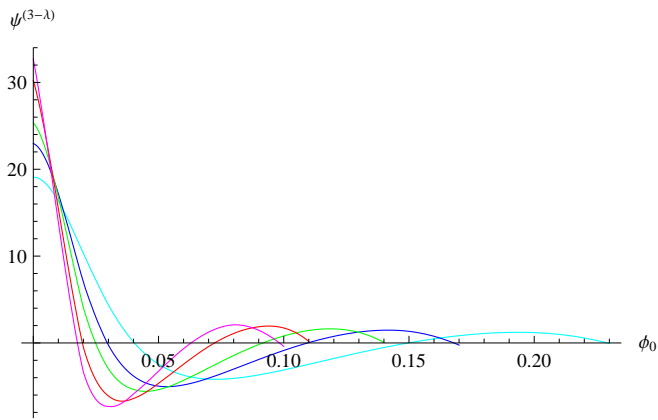}}}
\caption{
Dependence of the coefficient $\psi^{(3-\lambda)}$ on $\phi_0$ for
$m^2=-2$ ($\lambda=2$) $q=2$ for various staring points of
integration $\epsilon=1/500$ (cyan), $\epsilon=1/2000$ (blue),
$\epsilon=1/5000$ (green), $\epsilon=1/20000$ (red),
$\epsilon=1/50000$ (magenta). The smaller value of $\epsilon$ is
the closer zeros of $\psi^{(3-\lambda)}$ are
located.}\label{msnzm}
\end{figure}

Unfortunately, we were unable to construct a convergent procedure
of integration in this case. We used the following method. We
started the integration of (\ref{nlem}) from some point $\epsilon$
which is very close to the horizon $r=0$, substituting as an
initial condition the anzats (\ref{ans}). Then we decreased the
value of $\epsilon$ and compare the results. We did not observe
the convergence of the functions when decreasing $\epsilon$.
Namely, as $\epsilon$ approached zero the functions did not
approach a certain limit, showing significantly different
behavior. In the figure \ref{msnzm} we can see the dependence of
the coefficient $\psi^{(3-\lambda)}$ on the parameter $\phi_0$ for
various values of $\epsilon$. We checked that neither zeros of
$\psi^{(3-\lambda)}$ converge as $\epsilon\rightarrow0$ and,
therefore, we were not able to find the appropriate value of
$\phi_0$ for the solution that satisfies (\ref{normcond}). We have
checked also that addition of sub-dominant terms to the anzats
(\ref{ans}) does not remedy the situation.
\end{enumerate}

The $m=0$ case is free from the above problem of absence of
convergence and from here and on we shall consider only this case.

The equations (\ref{nlem}) have the two-parametric symmetry
\cite{Roberts}
\begin{equation}
r \rightarrow a r,\quad t \rightarrow \frac{b}{a} t, \quad g
\rightarrow a^2 g, \quad \phi \rightarrow \frac{a}{b}\phi, \quad
e^\chi \rightarrow b^2 e^\chi.
\end{equation}
One of these parameters allows us to fix
$\chi(r\rightarrow\infty)=0$. The other parameter re-scales the
parameters of the solutions and can be chosen so that the chemical
potential is unit.

In order to satisfy these two conditions, after the solution is
found we choose $$b=e^{-\chi(\infty)/2}, \quad a=b/\mu.$$


After this re-scaling the solution does not depend on the value of
$\chi_0$ and we can take $\chi_0=0$.


\section{Spectrum of the solutions}

\begin{figure}
\centerline{\resizebox{\linewidth}{!}{\includegraphics*{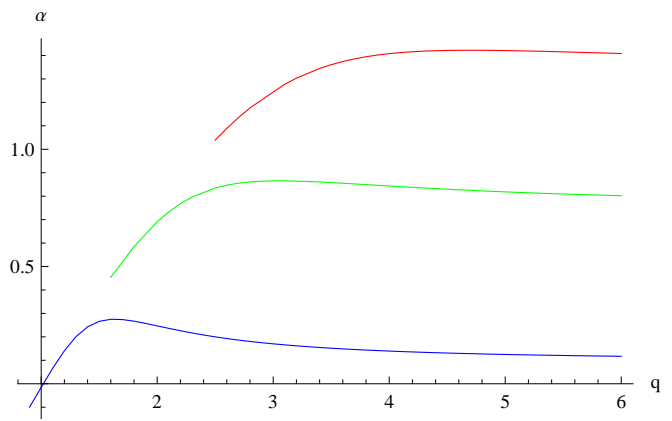}}}
\caption{
Three lowest solutions for $m=0$ given by $\alpha$ as functions of
$q$.}
\label{alphadep}
\end{figure}

\begin{figure*}
\centerline{\resizebox{\linewidth}{!}{\includegraphics*{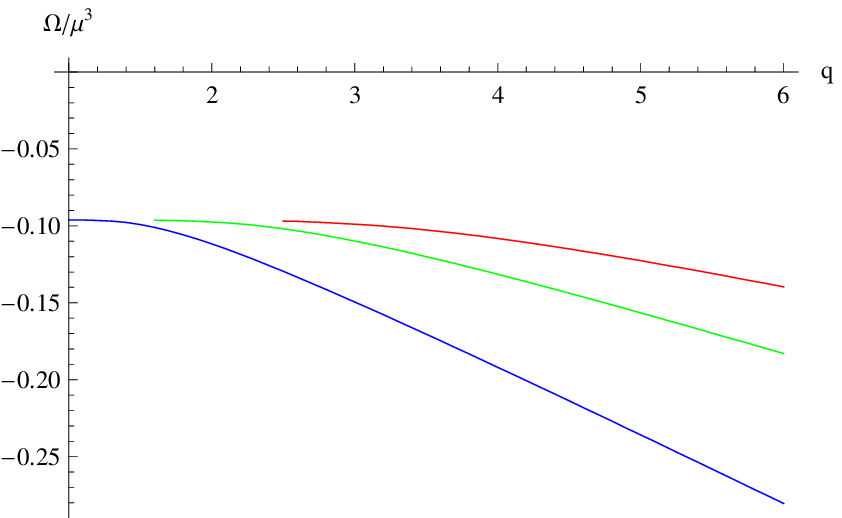}\includegraphics*{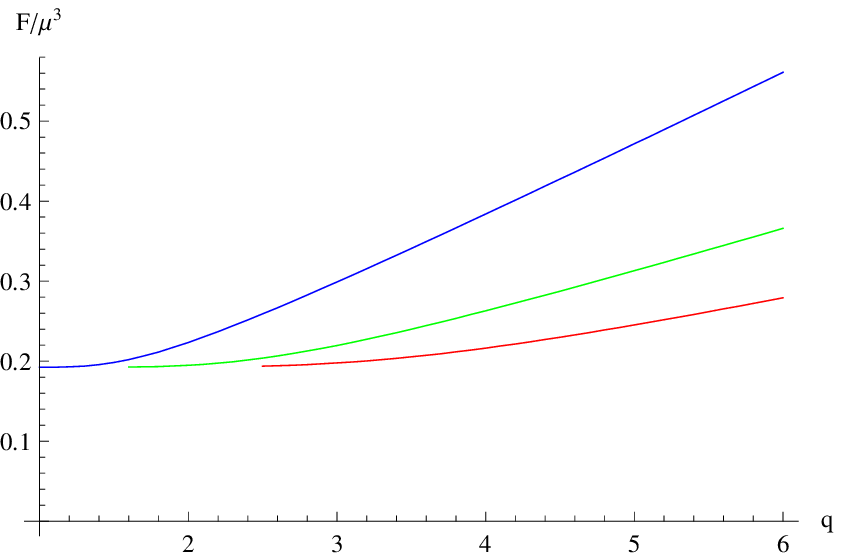}}}
\caption{
The densities of the grand canonical potential (left figure) and
the free energy (right figure) for the three lowest solutions for
$m=0$ as functions of $q$.}
\label{thermodynamics}
\end{figure*}

\begin{figure*}
\resizebox{0.5 \linewidth}{!}{\includegraphics*{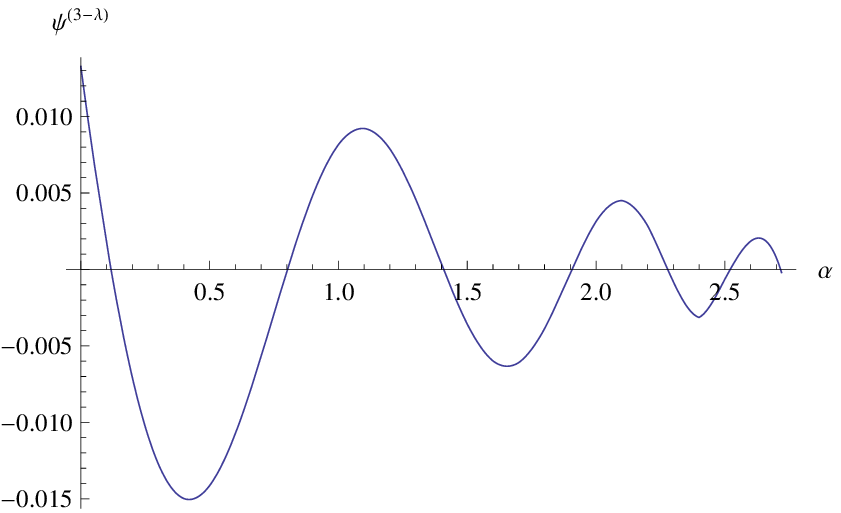}}\resizebox{0.5 \linewidth}{!}{\includegraphics*{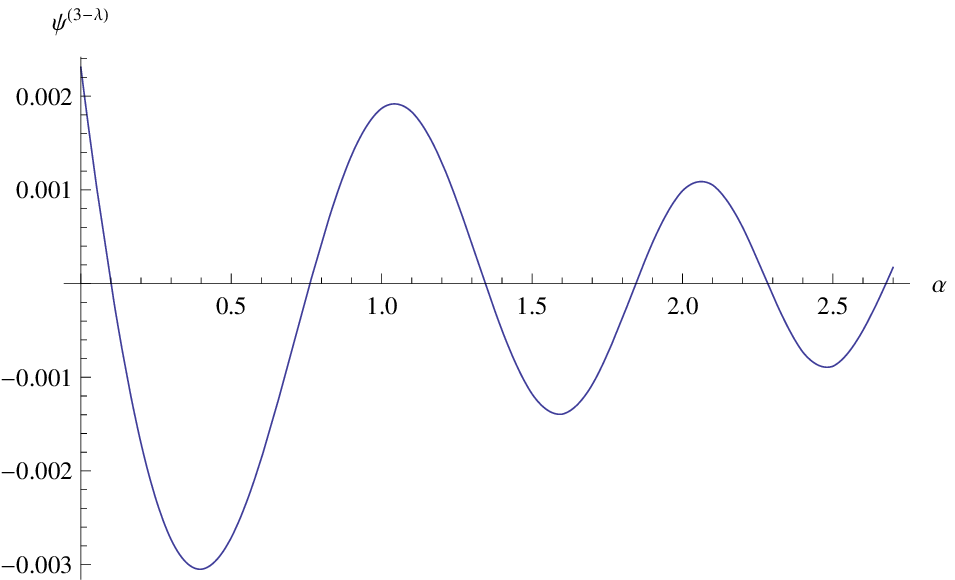}}
\caption{
Dependence of the coefficient $\psi^{(3-\lambda)}$ on $\alpha$ for $q=6$ (left) and $q=30$ (right).}
\label{Psidep}
\end{figure*}

In \cite{Roberts}, a value of $\alpha$ satisfying (\ref{normcond})
was found for $m=0$ as a function of $q$. It is interesting to
note, that for each fixed $q$ this value of $\alpha$ is not
unique. At least for large values of $q$, there is a discrete
spectrum of values of $\alpha$. Each of these value of $\alpha$
(under the same fixed $q$) corresponds to a
\emph{different solution} of (\ref{nlem}), that satisfies the
condition (\ref{normcond}). The dependence $q(\alpha)$ for the
first three solutions is shown on Fig.
\ref{alphadep}. We have checked that for all of the above three
curves the solutions are normalizable and reach their AdS
asymptotic at large distance. Near $r=0$, all three solutions obey
the same general anzats (\ref{ansatz}) though certainly with
different values of $\alpha$ for each $q$. Although we have
demonstrated only three solutions of the spectrum, it looks as if
there is an infinite spectrum of solutions with increasing values
of $\alpha$ for a fixed $q$.

On the figure \ref{Psidep} we see how the coefficient
$\psi^{(3-\lambda)}$ depends on $\alpha$ and on $q$: when $\alpha$
grows, the zeros of $\psi^{(3-\lambda)}$ become more and more
dense in $\alpha$, and when $q$ grows the he zeros of
$\psi^{(3-\lambda)}$ become more spaced. Therefore different
solutions (i.e. different lines $q(\alpha)$) lay closer to each
other for smaller $q$ and larger $\alpha$ making it difficult to
distinguish numerically different nearby solutions. That is why
the two upper curves do not continue on Fig. \ref{alphadep} to the
region where they probably coincide or lay very close to each
other: the numerical integration is not easy in that region as
there are probably many other solutions nearby. We believe however
that accurate numerical integration could allow to complete at
least a few upper curves until the minimal value of $q$.

\begin{figure}
\resizebox{\linewidth}{!}{\includegraphics*{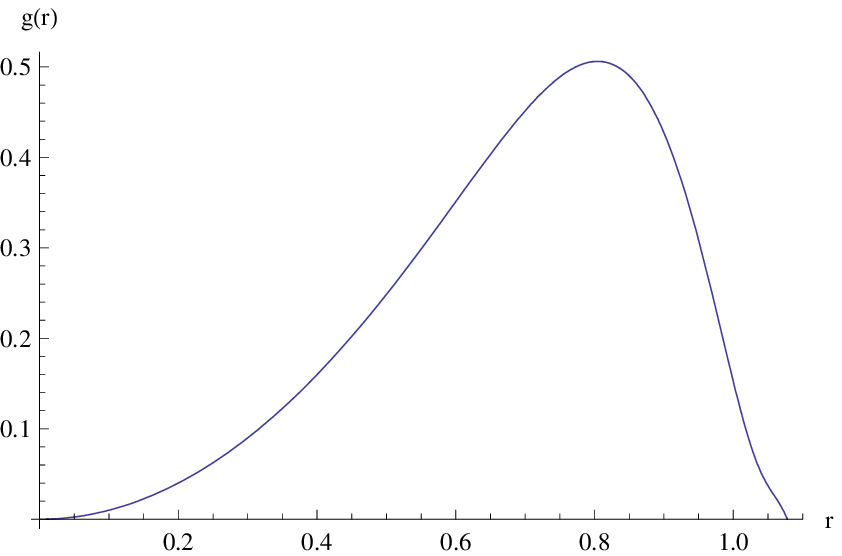}}
\caption{
Another singular point in the metric function $g(r)$ at $r>0$ when
$\alpha$ in the ansatz (\ref{ansatz}) is large ($m=0$, $q=6$,
$\alpha=2.8$).}
\label{singularg}
\end{figure}

For larger $\alpha$ we observe that another singular point appears
for $r>0$ (see Fig. \ref{singularg}).

The above found solutions correspond to \emph{lower} energy states
of the superconductor (see Fig. \ref{thermodynamics}). In order to
see this, we find the density of the grand canonical potential for
the corresponding states by fitting (\ref{gpfit}). Then, we can
calculate the density of the free energy, using the formula
\begin{equation}
F=\Omega+\mu\rho.
\end{equation}

\begin{figure*}
\resizebox{0.5\linewidth}{!}{\includegraphics*{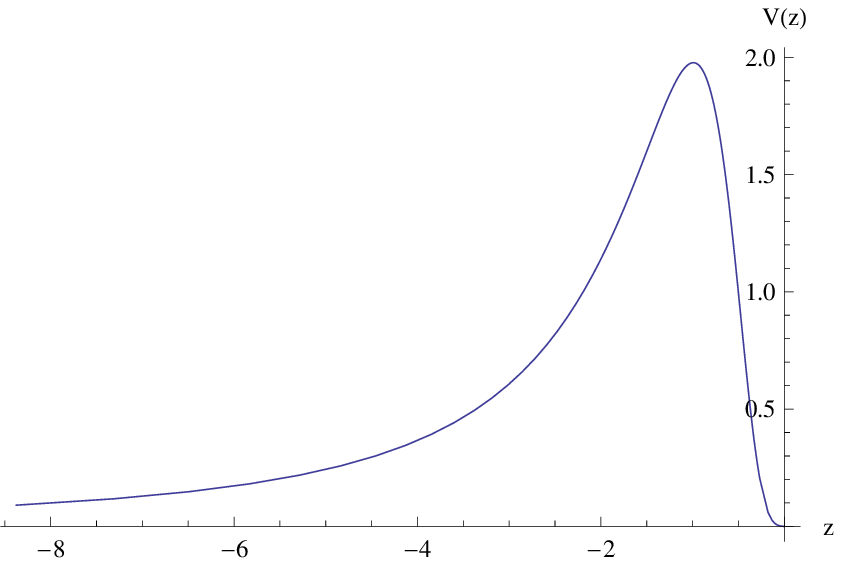}}\resizebox{0.5\linewidth}{!}{\includegraphics*{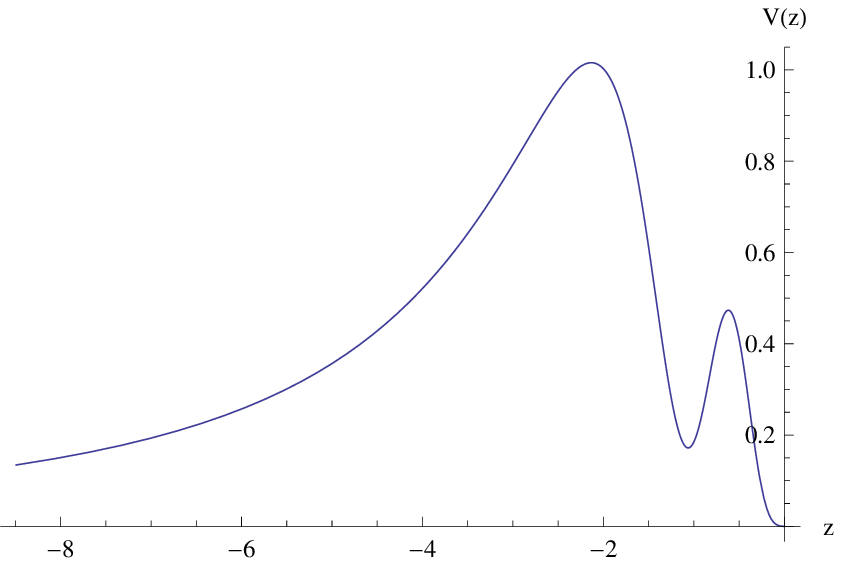}}
\resizebox{0.5\linewidth}{!}{\includegraphics*{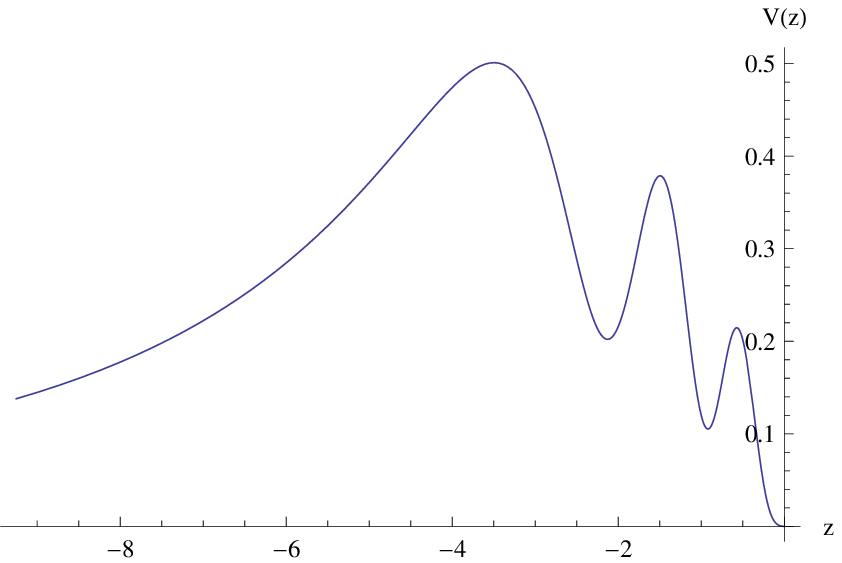}}\resizebox{0.5\linewidth}{!}{\includegraphics*{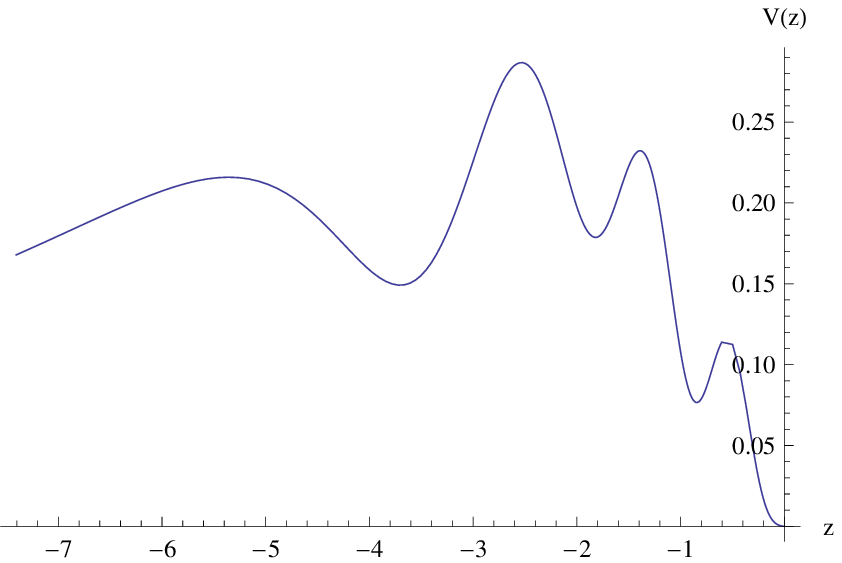}}
\caption{
The effective potentials as functions of the corresponding
tortoise coordinates for $m=0$, $q=6$ ($\lambda=3$) that for the
four smallest values of $\alpha$:} 
\begin{tabular}{c||r|r|r|r}
$\alpha$&$\rho/\mu^2$&$\psi^{(\lambda)}/\mu^\lambda$&$\Omega/\mu^3$&$F/\mu^3$\\
\hline
$0.117$&$0.84$&$0.79$&$-0.28$&$0.56$\\
$0.802$&$0.55$&$-0.58$&$-0.18$&$0.37$\\
$1.408$&$0.42$&$0.42$&$-0.14$&$0.28$\\
$1.907$&$0.35$&$-0.31$&$-0.12$&$0.24$\\
\end{tabular}
\label{7}
\end{figure*}

We found that $\Omega$ is larger for the state with larger
$\alpha$, but the free energy $F$ appears to be lower (see caption
for the figure \ref{7}).

In order to check our numerical calculations of the
thermodynamical potentials, we use the following relation between
the grand canonical potential and the free energy
\begin{equation}\label{traceless}
2\Omega=-F.
\end{equation}
This relation can be easily derived from (5.11) of
\cite{Hartnoll:2008kx} in the limit of zero temperature and magnetic
field. We find that (\ref{traceless}) is satisfied up to the
numerical precision.

On the figure \ref{alphadep} one can see the three smallest values
of $\alpha$ for which (\ref{normcond}) is satisfied. The smallest
$\alpha$ is the one found by Horowitz and Roberts. All the
potentials are positive definite, vanish at the horizon
($z=-\infty$) and at the spatial infinity ($z=0$). The potential
for the lowest value of $\alpha$ has one peak. The effective
potential for the $n$-th higher value of $\alpha$ has $n$ peaks
(Fig. \ref{7}). The larger value of $\alpha$ corresponds to the
state with larger charge density $\rho$larger absolute value of
the scalar hair $\psi^{(\lambda)}$ and lower density of the free
energy.

\section{Conductivity by the WKB method}

Assuming translational symmetry and stationary anzats in time, the linearized perturbation of the vector potential satisfies the wave-like equation  \cite{Hartnoll:2008kx}
\begin{equation}
A_x'' + \left(\frac{g'}{g} - \frac{\chi'}{2} \right) A_x' + \left(\left(\frac{\omega^2}{g^2} - \frac{\phi'^2}{g} \right) e^{\chi} - \frac{2 q^2
\psi^2}{g} \right) A_x  =  0 \,. \label{eq:ax}
\end{equation}
Using a new radial variable $dz = {e^{\chi/2}\over g} dr$,
at large $r$, $dz  = dr/r^2$, and we choose the constant of integration so that $ z = -1/r$.
The horizon is located at $z=-\infty$. Then (\ref{eq:ax}) has the wave-like form:
\begin{equation}\label{schr}
-A_{x,zz} + V(z) A_x = \omega^2 A_x,
\end{equation}
where the effective potential  \cite{Roberts}
\begin{equation}\label{potential}
V(z) = g[\phi_{,r}^2  + 2q^2 \psi^2  e^{-\chi}]
\end{equation}
As was shown in  \cite{Roberts} this effective potential always vanishes at the horizon. Since we consider only solutions which satisfy (\ref{normcond}), the potential also vanishes at the spatial infinity.

In terms of non-rescaled functions the potential and the tortoise coordinate are given by
\begin{eqnarray}
V(r) &=& \frac{a^2}{b^2}g\left(\phi_{,r}^2  + 2q^2 \psi^2  e^{-\chi}\right)=\frac{g}{\mu^2}\left(\phi_{,r}^2  + 2q^2 \psi^2  e^{-\chi}\right)\nonumber\\
dz&=&\frac{b}{a}\frac{e^{\chi/2}}{g}dr=\frac{e^{\chi/2}}{\mu g}dr.
\end{eqnarray}

According to the Horowitz-Roberts interpretation, the holographic
conductivity can be expressed in terms of the reflection
coefficients  \cite{Roberts} in the following way. In order to
solve (\ref{schr}) with the ingoing wave boundary conditions at $z
= -\infty$ we can extend the definition of the effective potential
to positive $z$ by setting $V=0$ for $z>0$ (the boundary of the
anti-de Sitter space (spatial infinity) is located at $z=0$).

Now an incoming wave from the right  will be partly transmitted
and partly reflected by the potential barrier. The transmitted
wave is purely ingoing at the horizon and the reflected wave
satisfies the scattering boundary conditions at $z \rightarrow
\infty$. Thus the scattering boundary conditions for $z>0$ are
\begin{equation}\label{BCs1}
A_x = e^{-i\omega z} + R e^{i\omega z}, \quad z \rightarrow +\infty,
\end{equation}
and at the event horizon
\begin{equation}\label{BCs2}
A_x = T e^{-i\omega z}, \quad z \rightarrow -\infty,
\end{equation}
where $R$ and $T$ are reflection and transmission coefficients.
Then one has
\begin{equation}
A_x(0) = 1+R, \quad A_{x,z}(0) = - i\omega(1-R).
\end{equation}
As shown in \cite{superconductor1}, if $ A_x = A_x^{(0)} + A_x^{(1)}/r$, and the conductivity is
\begin{equation}\label{oldcond}
\sigma(\omega) = -{i\over \omega} {A_x^{(1)}\over A_x^{(0)}}
\end{equation}
Since $ A_x^{(1)} = - A_{x,z} (0)$, so
\begin{equation}\label{cond}
\sigma(\omega) =  {1-R\over 1+R}
\end{equation}

The above boundary conditions (\ref{BCs1}), (\ref{BCs2}) are
nothing but the standard scattering boundary conditions for
finding the S-matrix. The effective potential has the distinctive
form of the potential barrier, so that the WKB approach \cite{WKB}
can be applied for finding $R$ and $\sigma$. Let us note, that as
the wave energy (or frequency) $\omega$ is real, the first order
WKB values for $R$ and $T$ will be real \cite{WKB} and
\begin{equation}\label{1}
T^2 + R^2 = 1.
\end{equation}
Next, we shall distinguish the two qualitatively different cases: first, when $\omega^2$ is much less then the maximum of the effective potential $\omega^2 \ll V_{0}$, and second when $\omega^2$ is of the same order that the maximum of the potential $\omega^2 \simeq V_{0}$ and can be either greater or smaller than the maximum. Strictly speaking, we should have to consider also the third case when $\omega^2$ is much larger than the maximum of the potential, but, as we shall see in most cases the reflection coefficient $R$ decreases too quickly with $\omega$, so that $\sigma$ reaches its maximal value (unity) even at moderate $\omega > V_{0}$.

For $\omega^2 \approx V_{0}$, we shall use the first order beyond the eikonal approximation WKB formula, developed by B. Schutz and C Will (see \cite{WKB}) for scattering around black holes
\begin{equation}\label{WKB1}
R = (1 + e^{- 2 i \pi (\nu + (1/2))})^{-\frac{1}{2}}, \quad \omega^2 \simeq V_{0},
\end{equation}
where
\begin{equation}
\nu + \frac{1}{2} = i \frac{(\omega^2 - V_{0})}{\sqrt{-2 V_{0}^{\prime \prime}}} + \Lambda_2 +\Lambda_3.
\end{equation}
Here $V_{0}^{\prime \prime}$ is the second derivative of the effective potential in its maximum, $ \Lambda_2$ and $ \Lambda_3$ are second and third WKB corrections which depend on up to 6th order derivatives of the effective potential at its maximum,

\begin{widetext}
\begin{eqnarray}
\Lambda_2&=&\frac{1}{(2Q^{''}_0)^{1/2}}\left\{\frac{1}{8}\left(\frac{Q^{(4)}_0}{Q^{''}_0}\right)
\left(\frac{1}{4}+N^2\right)-\frac{1}{288}\left(\frac{Q^{'''}_0}{Q^{''}_0}\right)^2
(7+60N^2)\right\},\\
\Lambda_3&=&\frac{N}{(2Q^{''}_0)^{1/2}}\bigg\{\frac{5}{6912}
\left(\frac{Q^{'''}_0}{Q^{''}_0}\right)^4 (77+188N^2)-
\frac{1}{384}\left(\frac{Q^{'''^2}_0Q^{(4)}_0}{Q^{''^3}_0}\right)
(51+100N^2)
+\frac{1}{2304}\left(\frac{Q^{(4)}_0}{Q^{''}_0}\right)^2(67+68N^2)
\nonumber\\&+&\frac{1}{288}
\left(\frac{Q^{'''}_0Q^{(5)}_0}{Q^{''^2}_0}\right)(19+28N^2)-\frac{1}{288}
\left(\frac{Q^{(6)}_0}{Q^{''}_0}\right)(5+4N^2)\bigg\},
\end{eqnarray}
\end{widetext}
and
$$
N=\nu+\frac{1}{2},\quad
Q^{(n)}_0=\frac{d^nQ}{dr^n_*}\bigg|_{r_*=r_*(r_{max})},
\quad Q
\equiv \omega^2 - V.
$$

The above formula was extended up to the 6th WKB order in \cite{WKBorders} and applied to a number of problems of scattering around black holes (see for instance \cite{WKBuse} and references therein). Mainly it was used for finding the so-called quasinormal modes of black holes, which imply special boundary conditions, so that $\nu$ becomes integer in that case. For arbitrary $\nu$ and each given $\omega$ the above WKB formula works for problems with the standard scattering boundary conditions.
We shall look for higher WKB orders in order to have the idea of possible order of the error in the obtained results. Though the WKB series converges only asymptotically, in many cases, quite unexpectedly, WKB values have region of relative convergence in orders.

The case of small frequencies is well described by the well-known formula
\begin{equation}
T = e^{- \int_{z_1}^{z_2} d z \sqrt{V(z) - \omega^2} }, \quad \omega^2 \ll V_{0}
\end{equation}
i.e. the at small frequency the transmission is exponentially suppressed. Here $z_1$ and $z_2$ are the turning points for which $V(z) = \omega^2$.
The reflection coefficient follows from (\ref{1})
\begin{equation}\label{WKB2}
R = \sqrt{1 - e^{-2 \int_{z_1}^{z_2} d z \sqrt{V(z) - \omega^2} }}, \quad \omega^2 \ll V_{0}
\end{equation}

\begin{figure}
\centerline{\resizebox{\linewidth}{!}{\includegraphics*{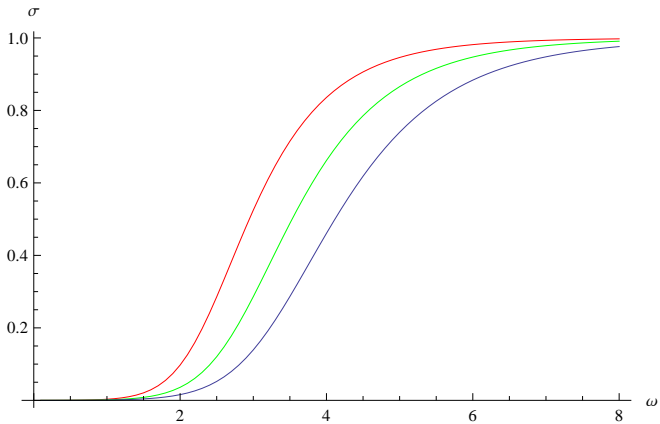}}}
\caption{
The conductivity found by the WKB formula 
for $q=10$ (top, red), $q=12$ (green), $q=14$ (bottom, blue) as a function of frequency $\omega$ for $m=0$. }
\label{F1}
\end{figure}

\begin{figure*}
\resizebox{0.5 \linewidth}{!}{\includegraphics*{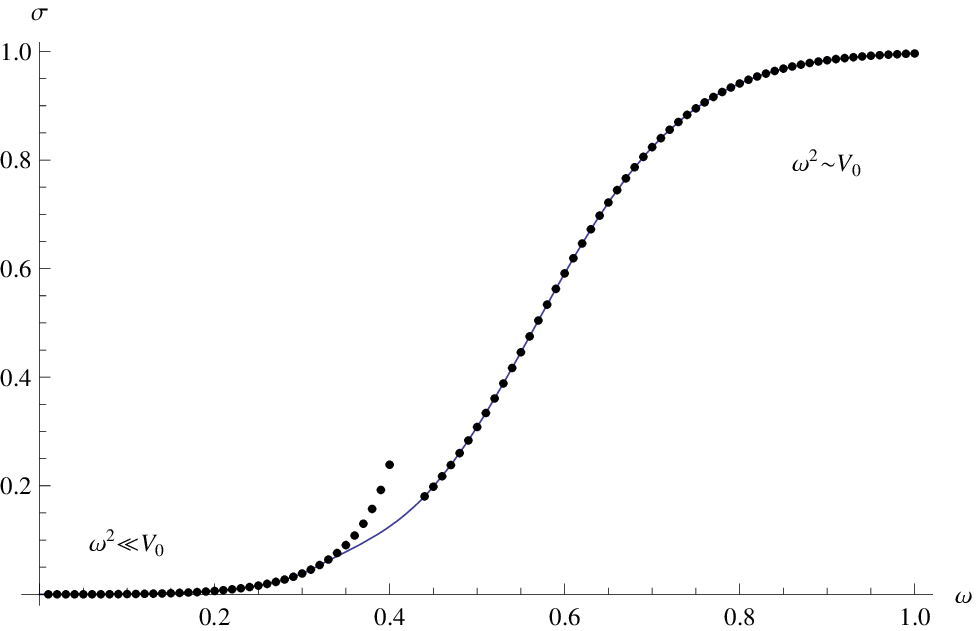}}\resizebox{0.5 \linewidth}{!}{\includegraphics*{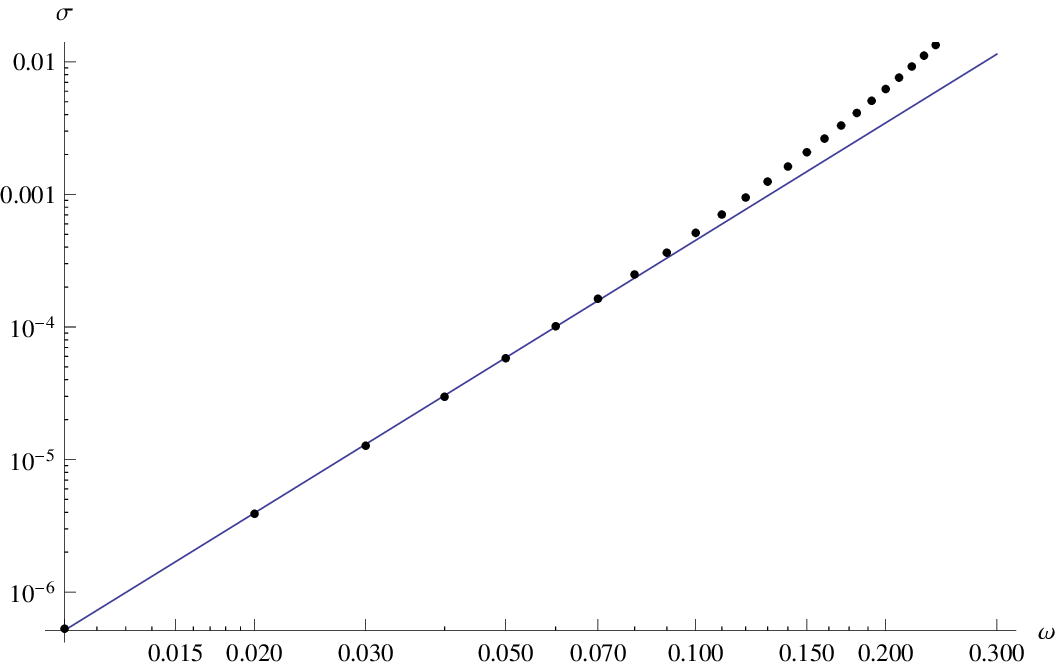}}
\caption{
The conductivity for $m=0$, $q=1$ as a function of frequency
$\omega$ ($\delta\approx3.0$). On the left figure the solid line
corresponds to the interpolation between the two WKB
approximations (\ref{WKB1}) and (\ref{WKB2}). On the right figure
the solid line corresponds to the fit of the numerical data for
small values of $\omega$. }
\label{F2}
\end{figure*}

\begin{figure*}
\resizebox{0.5 \linewidth}{!}{\includegraphics*{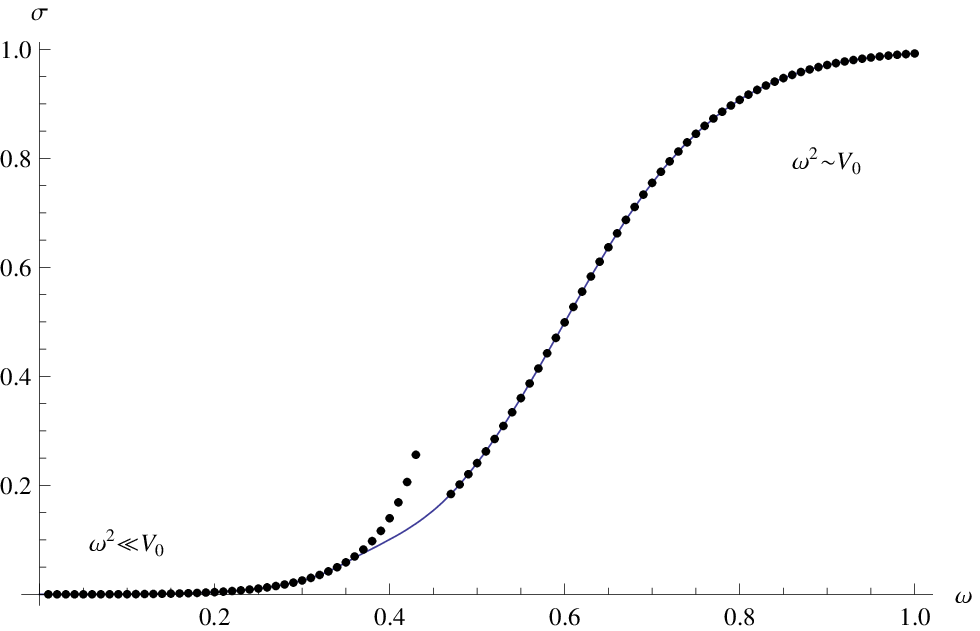}}\resizebox{0.5 \linewidth}{!}{\includegraphics*{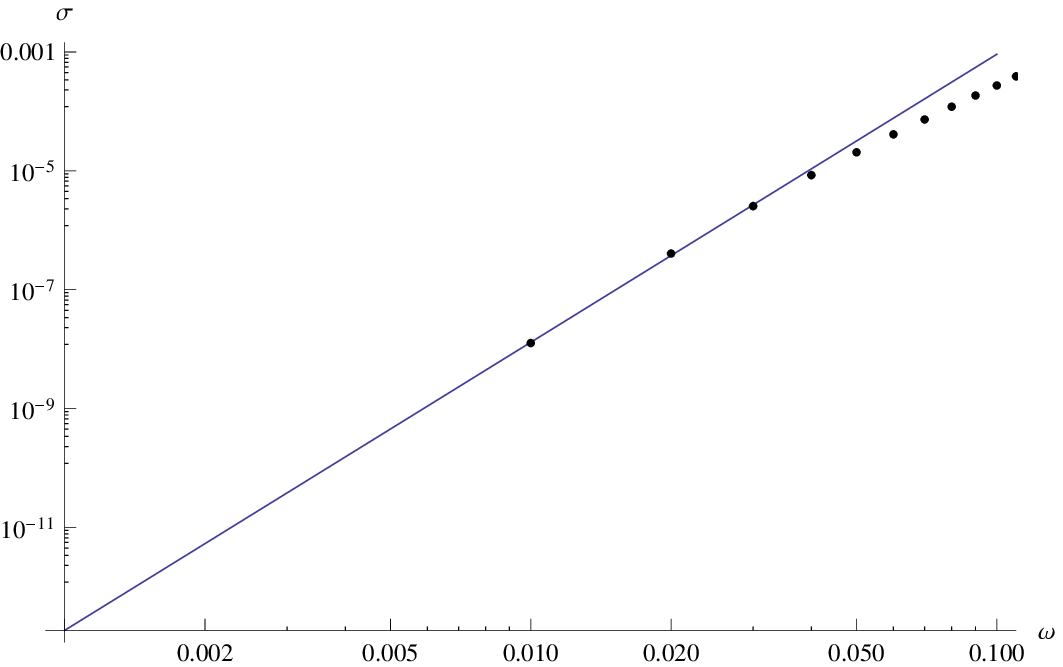}}
\caption{
The conductivity for $m=0$, $q=1.6$ as a function of frequency
$\omega$ ($\delta\approx4.8$). On the left figure the solid line
corresponds to the interpolation between the two WKB
approximations (\ref{WKB1}) and (\ref{WKB2}). On the right figure
the solid line corresponds to the fit of the numerical data for
small values of $\omega$.}
\label{F3}
\end{figure*}

\begin{figure*}
\resizebox{0.5 \linewidth}{!}{\includegraphics*{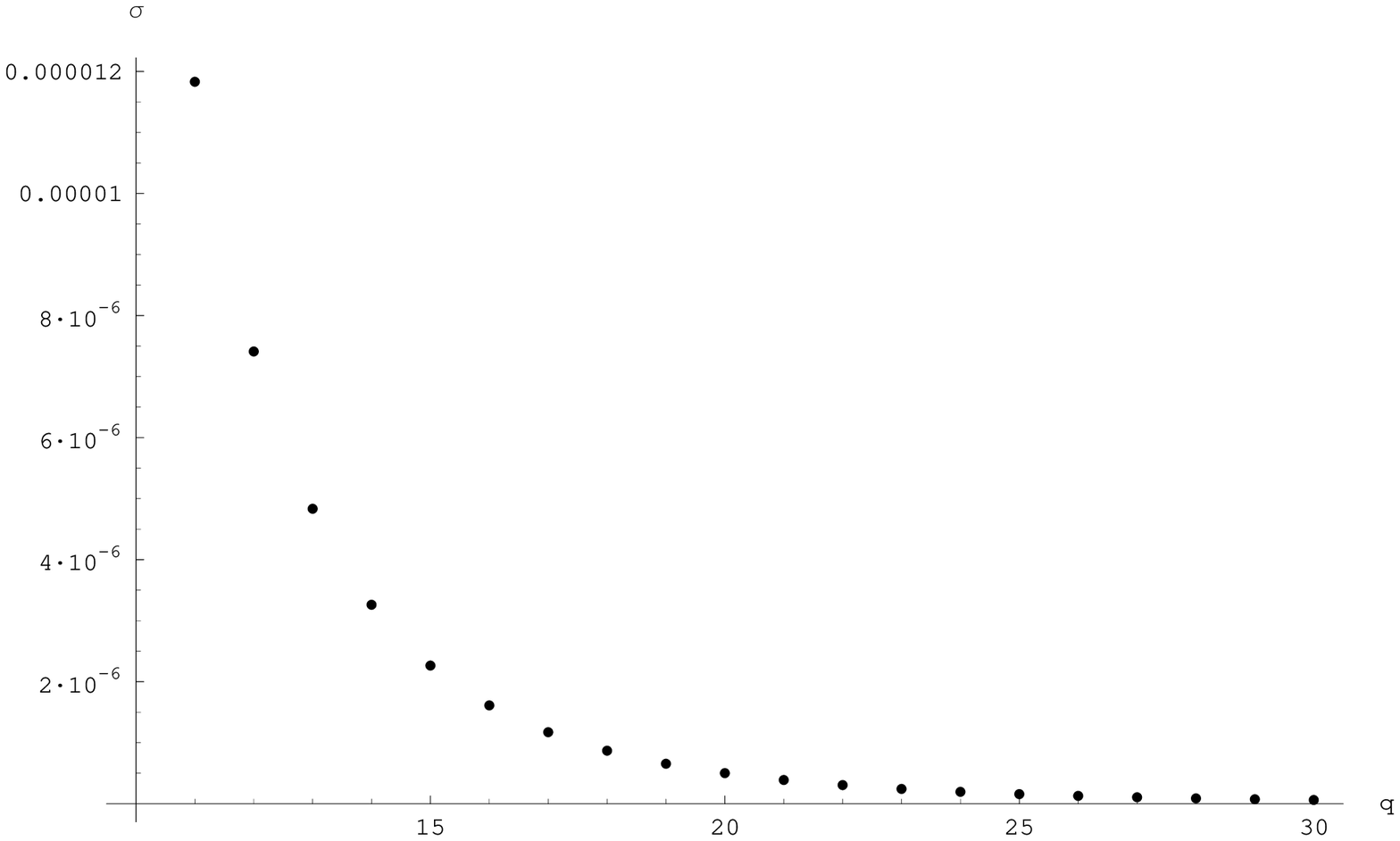}},\resizebox{0.5 \linewidth}{!}{\includegraphics*{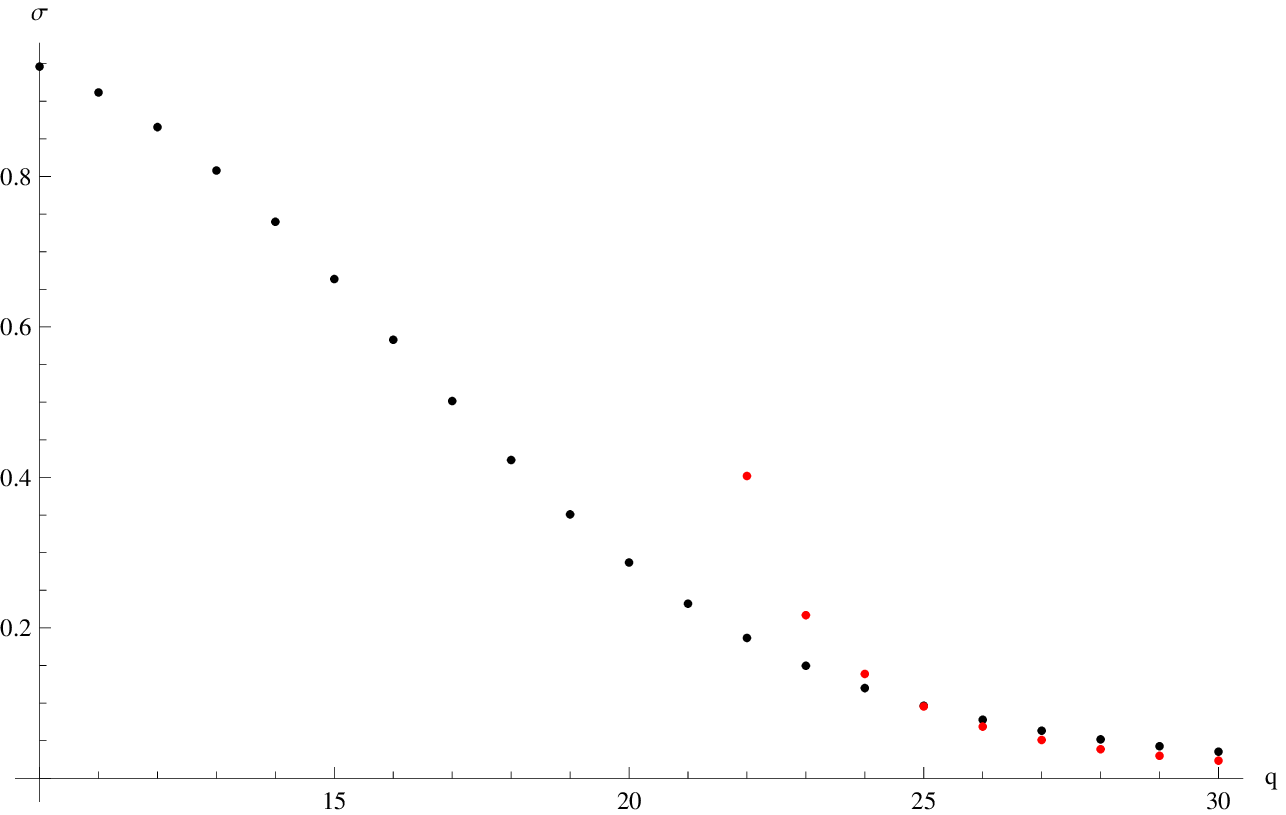}}
\caption{
WKB conductivity for $m=0$ and for various values $q$ $\omega =1/2$ (left) and $\omega =5$ (right). As $q$ grows, the maximum of the effective potential grows, so that a fixed $\omega$ moves down from the peak, approaching the regime $\omega^2 \ll V_{0}$ (red dots).}
\label{F35}
\end{figure*}

At small frequencies the reflection coefficient is close to $1$,
i.e. almost all energy is reflected by the potential. Then $R$
decreases with the increasing of $\omega$, and, for sufficiently
large $\omega$, usually seemingly larger than $V_{0}$ or about it,
the reflection coefficient is close to zero. This means that
according to (\ref{cond}), the conductivity changes from zero at
small frequencies until $1$ at large frequencies. This kind of
behavior we can see on Fig. \ref{F1}, where the conductivity was
obtained by using and the expression (\ref{cond}) and the WKB
formula (\ref{WKB2}) (for $\omega\leq 0.4 $) and (\ref{WKB1}) (for
$\omega\geq 0.4$). There one can see that as $\omega^2$ approaches
the peak of the potential barrier (which is located at $r \approx
0.18 $), the accuracy of the formula (\ref{WKB2}) diminishes and
(\ref{WKB1}) becomes a better approximation. Though a good
confirmation of consistency of the both approximations
(\ref{WKB2}) and (\ref{WKB1}) is the possibility of "smooth
matching" of both data (see Fig. \ref{F2}, \ref{F3}) if neglecting
small intermediate region of $\omega$, where both approximation
have marginal accuracy (For Fig. 2, this intermediate region is
$0.3 \leq \omega \leq 0.4$). In some range of parameters, such as
the one shown on Fig. 3 (right), there is no such intermediate
region that should be neglected but, even better, both regimes
(\ref{WKB2}) and (\ref{WKB1}) overlap, giving almost the same
values for some range of large values of $q$.

Let us note here two important technical points. First is that
when using formula (\ref{WKB2}) one needs the higher order
derivatives of the effective potential which is unknown in
analytical form, but is given only numerically. It would be a
rough method to approximate the effective potential by some
interpolating analytical function and then to take derivatives of
it: each derivative would bring additional numerical error to the
calculations. Instead we used the field equations (\ref{nlem}) and
have taken all necessary derivatives from (\ref{nlem})  and by
taking the corresponding derivatives of the wave and metric
functions $\phi$, $\psi$, $g$, etc..

Another important moment is the accuracy of the used WKB
technique. The existence of the ``common region'' where both
formulas produce the same result says that for large values of $q$
the WKB formulas work very well. The analysis of the higher order
corrections indeed shows that for large $q$ (and $\omega^2 \simeq
V_{0}$) the WKB series shows convergence in a few first orders: An
example is $q=10$, $\omega=2.2$, $R = 0.778$ for the first WKB
order, $R =0.722$ for second WKB order, and $R = 0.725$ for the
third order. This gives estimated error of less than one percent.
There is no such good convergence for small values of $q$,
therefore for $\omega^2 \simeq V_{0}$ we have used here the WKB
formula of the first order for small $q$, and 3th order formula
for large $q$.

At small frequencies we have obtained a close, though not
coinciding, numerical result to the formula (3.21) of
\cite{Roberts}
\begin{equation}\label{asymptot}
\sigma = \left(\frac{\omega}{\omega_0}\right)^{\delta}, \quad \delta = \sqrt{4 V_{0} +1} - 1.
\end{equation}
Thus for $q=1.6$, we obtained by WKB $\delta \approx 4.8$
($\omega_0=0.45$), what is close to $\sqrt{4 V_{0} +1} -
1\approx4.55$, while  for $q=1$ WKB gives $\delta \approx 3.0$
($\omega_0\approx1.3$) and  $\sqrt{4 V_{0} +1} - 1\approx3.97$.
The WKB correction to the (\ref{asymptot}), as it can be seen from
the above data, may be quite big and about 25 per cents.

The formula (\ref{asymptot}) used for its derivation rough
matching of the left and right dominant asymptotics. Thus it is
expected to be less accurate than the WKB method we are using
here.

\section{Conclusions}

We have found the WKB values of conductivity for the
Horowitz-Roberts model of the zero-temperature superconductor for
$m^2 =0$ case. Dependence of conductivity on parameters of the
theory such as the charge density $\rho$ and the frequency
$\omega$ is investigated. WKB data for conductivity confirms the
qualitative arguments that $\sigma$ does not reach zero even at
zero temperature, in agreement with \cite{Roberts}. By the WKB
calculations  we have confirmed the analytic relation derived in
\cite{Roberts} for the $\omega$-dependence of $\sigma$ at small
frequencies and calculated the pre-factor for this relation for
various $q$. The used here third order WKB formula which has very
good accuracy for large values of $q$ (and moderate $\omega$),
showing convergence in orders with an estimated error of around
fractions of one percent. In addition, we have found the set of
other solutions which describe the superconductor at zero
temperature in the states with higher grand canonical potential
and lower energy.

Our paper may be improved in a number of ways. First of all, the
conductivity values could be obtained with better accuracy, if one
uses the numerical shooting, which is known to work well for
asymptotically AdS space-times \cite{myRNADS}. The conductivity of
higher grand potential states with $m=0$ cannot be obtained by WKB
formula we used, because the effective potentials for higher
states have a number of local maximums. Thus accurate shooting
approach would allow also for complete analysis of conductivities
of these states. Finally, the case of non-vanishing mass of the
scalar field $m$ probably requires some other and more
sophisticated procedure of integration or a different anzats near
the horizon from that suggested in \cite{Roberts}.

\section*{Acknowledgments}
At the initial stage of this work R. A. K. was supported by
\emph{the Japan Society for the Promotion of Science} (JSPS),
Japan and at the final stage by \emph{the Alexander von Humboldt
foundation} (AvH), Germany. \mbox{A. Z.} was supported by
\emph{Funda\c{c}\~ao de Amparo \`a Pesquisa do Estado de S\~ao
Paulo (FAPESP)}, Brazil. A. Z. also acknowledges the hospitality
of the University of Guadalajara, M\'exico, where a part of this
work was done.

\end{document}